# A MATLAB Code for Three Dimensional Linear Elastostatics using Constant Boundary Elements

Kirana Kumara P
Centre for Product Design and Manufacturing
Indian Institute of Science
Bangalore, India
e-mail: kiranakumarap@gmail.com

*Abstract*—**Present work presents a code written in the very simple programming language MATLAB, for three dimensional linear elastostatics, using constant boundary elements. The code, in full or in part, is not a translation or a copy of any of the existing codes. Present paper explains how the code is written, and lists all the formulae used. Code is verified by using the code to solve a simple problem which has the well known approximate analytical solution. Of course, present work does not make any contribution to research on boundary elements, in terms of theory. But the work is justified by the fact that, to the best of author's knowledge, as of now, one cannot find an open access MATLAB code for three dimensional linear elastostatics using constant boundary elements. Author hopes this paper to be of help to beginners who wish to understand how a simple but complete boundary element code works, so that they can build upon and modify the present open access code to solve complex engineering problems quickly and easily. The code is available online for open access (as supplementary file for the present paper), and may be downloaded from the website for the present journal.**

## I. INTRODUCTION

Ready availability of computer codes encourages the use of any numerical technique. Reference [1], in its last paragraph of the section "Conclusion", opines that boundary elements are a good idea in principle but not in practice because of the lack of proper software. This opinion remains true even today.

There are some open source BEM libraries. Helsinki BEM library [2] is a MATLAB source code library for problems that obey the Laplace or Poisson equation. The web source [3] contains codes that are specifically useful for solving acoustics problems. The source also contains codes for solving Laplace problems and Helmholtz problems. Book [4] gives FORTRAN codes for Laplace's equation and Helmholtz equation, in two and three dimensions. The codes can be freely downloaded from [5], website for the book. A BEM code for two dimensional (2D) pulsating cylinders is available from [6].

Fast Multipole Boundary Element Method (FastBEM) software is available from [7]. Software for three dimensional (3D) elasticity is also available here. Still, source codes are not available here.

The website [8] for the book [9] contains many programs in Fortran. It contains programs for 3D elasticity also. But the website tells that the programs supplied there are for use by purchasers of the book only. Here, the Fortran program for 3D elasticity is not written as a single program; the main program calls different modules to perform different tasks. It is difficult to fully understand programs, without referring to the book.

But, although three dimensional elasticity is such an important area, apart from the codes which might be available in the websites (e.g., the website [8]) that are companions to some non open access books, one cannot find an *open access source code* in any of the programming languages, although author of this paper could find a program on two dimensional elasticity in a file sharing system. Extension of a code on two dimensional elasticity into three dimensions is not very trivial and one needs some new formulae also. Also, file sharing systems sometimes delete some of the hosted files. Since codes are readily available for potential problems in three dimensions, it may be possible to cast a three dimensional elasticity problem as a three dimensional potential problem using potential representations like Papkovich-Neuber representation, but this is not the standard way an elasticity problem is solved using boundary elements. Hence, author of the present work thought of writing the present code on three dimensional elasticity and making it available for open access, through the present paper in the present open access journal. Present paper does not explain the theory behind boundary elements in detail. Aim is to list all the formulae that are needed to write the code, and explain how these formulae are assembled to produce a working code. The present paper is helpful to understand the working of the present code.

Present work does not aim to provide source code for whole of three dimensional elasticity. It provides a MATLAB code only for the most basic form of three dimensional elasticity, i.e., three dimensional linear elastostatics. Although very basic, three dimensional linear elastostatics has wide applications in product design and structural design. Purpose of selecting MATLAB is that it is very easy to learn, and people who do not know the language also can follow the logic of the code. Using Parallel Computing Toolbox, now a MATLAB code can very easily be parallelized to run on multiple CPUs/GPUs. Code can be precompiled to increase speed. While solving complex real-world problems, a MATLAB code can readily interact with already developed subroutines in other languages like C/C++ and Fortran, using 'External Interfaces' feature of MATLAB. With little modification, a MATLAB code may be executed in one of open source and free equivalents of MATLAB such as GNU Octave, FreeMat and Scilab. Although very simple and very basic, the present code is not a subroutine but a complete program. Also, the present code does not contain any subroutines. Even input data has to be entered in the code itself. In terms of theory, present work does not make any contribution to research on boundary elements. All theory behind the present code, including all formulae, is taken from [9] and [4]. But the present code, in full or in part, is not a translation or a copy of any of the existing codes.

The present work may also be useful as an educational aid to learn the basics of the boundary element method as applied to 3D linear elastostatics especially since it uses the most basic form of 3D elasticity, i.e., 3D linear elastostatics, and the most basic form of elements, constant elements. It may be noted that [10] presents a way of implementing the boundary element method using MATLAB, including details on coding, but for solving the Laplace's equation only. With detailed explanation of the theory, a MATLAB code for two dimensional Laplace's equation is presented in [11]; it makes use of constant elements.

Present paper is organized as follows. Next section describes the theory that is essential to develop the code. The subsequent section explains the code. The section that follows illustrates the use of the code to solve a well known simple problem which has a well known solution, and thus verifies the code.

## II. THEORY

Only theory that is essential to understand the present code is explained here. One can refer to [4] and [9] for further details.

From the Appendix of [9], for static elasticity, in indicial notation, the displacement $u_i$ at an internal point $P$, in the absence of initial stresses and strains, is given by

$$u_i(p) = \int_S U_{ij}(P,Q) t_j(Q) dS - \int_S T_{ij}(P,Q) u_j(Q) dS$$

(1)

where $u_i, t_i$ (or $u_j, t_j$) are the displacements and tractions

$U_{ij}(P,Q), T_{ij}(P,Q)$ are called the fundamental solutions

$P$ is called the source point and $Q$ is called the field point

$S$ is the surface (for 3D problems) which represents the geometry

For 3D problems, the fundamental solutions are given by

$$U_{ij}(P,Q) = C(C_1 \ln \frac{1}{r} \delta_{ij} + r_{,i} r_{,j})$$

(2)

$$T_{ij}(P,Q) = \frac{-C_2}{r^n} \left[ \begin{array}{c} (C_3 \delta_{ij} + (n+1) r_{,i} r_{,j}) \cos\theta \\ -C_3 (1 - \delta_{ij})(n_j r_{,i} - n_i r_{,j}) \end{array} \right]$$

(3)

In (2) and (3), $r$ is the distance between $P$ and $Q$, and $n_i$ and $n_j$ are the outward normals. The derivative of $r$ with respect to the Cartesian axis $i$ is denoted as $r_{,i}$ and the derivative of $r$ with respect to the Cartesian axis $j$ is denoted as $r_{,j}$. The term $\cos\theta$ is given by

$$\cos\theta = \frac{1}{r} \vec{r} \cdot \vec{n}$$

(4)

$\delta_{ij}$ is given by

$$\delta_{ij} = \begin{Bmatrix} 1 \, for(i = j) \\ 0 \, for(i \neq j) \end{Bmatrix}$$

(5)

The values of the constants are given by

$$n = 2$$

$$C = \frac{1}{16\pi G(1-v)}$$

$$C_1 = 3 - 4v$$

$$C_2 = \frac{1}{8\pi(1-v)}$$

$$C_3 = 1 - 2v$$

(6)

where $\nu$ is the Poisson's ratio.

The shear modulus $G$ is given by

$$G = \frac{E}{2(1+\nu)}$$

(7)

where $E$ is the modulus of elasticity.

In the present work, a 3D solid is represented by 3D boundary triangles, i.e., 3D triangular surface mesh. $T$ is the total number of triangles which together represent the 3D solid; hence, the total number of elements is equal to $T$. Let $S^m$ be the surface of the element with element number $m$. Here, since constant elements are used, over each of the elements, displacements and tractions are assumed constant. For each of the elements, either displacement or traction is known, the other being an unknown that has to be calculated. In this work, solution is sought only on the boundary. For a point $P$ on the boundary of a solid, if $P$ is located inside a smooth region of the boundary, (1) can be reduced to the following three equations, i.e., (8), (9) and (10).

$$\frac{1}{2}u_x(P_e) = \sum_{m=1}^{T}[A1 + A2 + A3 - B1 - B2 - B3]$$

(8)

$$\frac{1}{2}u_y(P_e) = \sum_{m=1}^{T}[A4 + A5 + A6 - B4 - B5 - B6]$$

(9)

$$\frac{1}{2}u_z(P_e) = \sum_{m=1}^{T}[A7 + A8 + A9 - B7 - B8 - B9]$$

(10)

where $A1 = t_x(Q_m)\int_{S^m} U_{xx}(P_e, Q_m)dS^m$

$A2 = t_y(Q_m)\int_{S^m} U_{xy}(P_e, Q_m)dS^m$

$A3 = t_z(Q_m)\int_{S^m} U_{xz}(P_e, Q_m)dS^m$

$B1 = u_x(Q_m)\int_{S^m} T_{xx}(P_e, Q_m)dS^m$

$B2 = u_y(Q_m)\int_{S^m} T_{xy}(P_e, Q_m)dS^m$

$B3 = u_z(Q_m)\int_{S^m} T_{xz}(P_e, Q_m)dS^m$

$A4 = t_x(Q_m)\int_{S^m} U_{yx}(P_e, Q_m)dS^m$

$A5 = t_y(Q_m)\int_{S^m} U_{yy}(P_e, Q_m)dS^m$

$A6 = t_z(Q_m)\int_{S^m} U_{yz}(P_e, Q_m)dS^m$

$B4 = u_x(Q_m)\int_{S^m} T_{yx}(P_e, Q_m)dS^m$

$B5 = u_y(Q_m)\int_{S^m} T_{yy}(P_e, Q_m)dS^m$

$B6 = u_z(Q_m)\int_{S^m} T_{yz}(P_e, Q_m)dS^m$

$A7 = t_x(Q_m)\int_{S^m} U_{zx}(P_e, Q_m)dS^m$

$A8 = t_y(Q_m)\int_{S^m} U_{zy}(P_e, Q_m)dS^m$

$A9 = t_z(Q_m)\int_{S^m} U_{zz}(P_e, Q_m)dS^m$

$B7 = u_x(Q_m)\int_{S^m} T_{zx}(P_e, Q_m)dS^m$

$B8 = u_y(Q_m)\int_{S^m} T_{zy}(P_e, Q_m)dS^m$

$B9 = u_z(Q_m)\int_{S^m} T_{zz}(P_e, Q_m)dS^m$

To be clearer, equations (8)-(10) may also be written in the expanded form given by the following three equations, i.e., (11), (12) and (13).

$$\frac{1}{2}u_x(P_e) = [A11 + A21 + A31 - B11 - B21 - B31]$$

$$+ [A12 + A22 + A32 - B12 - B22 - B32] + \ldots$$

$$\ldots + [A1m + A2m + A3m - B1m - B2m - B3m] + \ldots$$

$$\ldots + [A1T + A2T + A3T - B1T - B2T - B3T]$$

(11)

$$\frac{1}{2}u_y(P_e) = [A41 + A51 + A61 - B41 - B51 - B61]$$

$$+ [A42 + A52 + A62 - B42 - B52 - B62] + \ldots$$

$$\ldots + [A4m + A5m + A6m - B4m - B5m - B6m] + \ldots$$

$$\ldots + [A4T + A5T + A6T - B4T - B5T - B6T]$$

(12)

$$\frac{1}{2}u_z(P_e) = [A71 + A81 + A91 - B71 - B81 - B91]$$

$$+ [A72 + A82 + A92 - B72 - B82 - B92] + \ldots$$

$$\ldots + [A7m + A8m + A9m - B7m - B8m - B9m] + \ldots$$

$$\ldots + [A7T + A8T + A9T - B7T - B8T - B9T]$$

(13)

where 
$$A1m = t_x(Q_m)\int_{S^m} U_{xx}(P_e, Q_m)dS^m$$

$$A2m = t_y(Q_m)\int_{S^m} U_{xy}(P_e, Q_m)dS^m$$

$$A3m = t_z(Q_m)\int_{S^m} U_{xz}(P_e, Q_m)dS^m$$

$$B1m = u_x(Q_m)\int_{S^m} T_{xx}(P_e, Q_m)dS^m$$

$$B2m = u_y(Q_m)\int_{S^m} T_{xy}(P_e, Q_m)dS^m$$

$$B3m = u_z(Q_m)\int_{S^m} T_{xz}(P_e, Q_m)dS^m$$

$$A4m = t_x(Q_m)\int_{S^m} U_{yx}(P_e, Q_m)dS^m$$

$$A5m = t_y(Q_m)\int_{S^m} U_{yy}(P_e, Q_m)dS^m$$

$$A6m = t_z(Q_m)\int_{S^m} U_{yz}(P_e, Q_m)dS^m$$

$$B4m = u_x(Q_m)\int_{S^m} T_{yx}(P_e, Q_m)dS^m$$

$$B5m = u_y(Q_m)\int_{S^m} T_{yy}(P_e, Q_m)dS^m$$

$$B6m = u_z(Q_m)\int_{S^m} T_{yz}(P_e, Q_m)dS^m$$

$$A7m = t_x(Q_m)\int_{S^m} U_{zx}(P_e, Q_m)dS^m$$

$$A8m = t_y(Q_m)\int_{S^m} U_{zy}(P_e, Q_m)dS^m$$

$$A9m = t_z(Q_m)\int_{S^m} U_{zz}(P_e, Q_m)dS^m$$

$$B7m = u_x(Q_m)\int_{S^m} T_{zx}(P_e, Q_m)dS^m$$

$$B8m = u_y(Q_m)\int_{S^m} T_{zy}(P_e, Q_m)dS^m$$

$$B9m = u_z(Q_m)\int_{S^m} T_{zz}(P_e, Q_m)dS^m$$

where $m$ takes values from 1 to $T$.

Equations (8)-(10) (or equations (11)-(13)) are the basic equations upon which the present code is developed. Since displacements and tractions are constants over each of the elements, for each of the elements, displacements and tractions are considered only for just one chosen point inside each element. $P_e$ and $Q_m$ refer to these points; here, the subscripts $e$ or $m$ in $P_e$ or $Q_m$ refer to the element number. The subscript $e$ in $P_e$ varies from 1 to $T$ which is the total number of elements. Further, $m = e$ implies that $P_e = Q_m$. Hence, if a solid is discretized by $T$ boundary elements, equation (8)-(10) (or equation (11)-(13)) give rise to a set of coupled $3T$ linear algebraic equations in $3T$ unknowns. Unknowns are either displacements ($u_x$, $u_y$ or $u_z$ at $P_e$ or $Q_m$) or tractions ($t_x$, $t_y$ or $t_z$ at $Q_m$ (or $P_e$ when $e = m$)). For elements with prescribed displacements ($u_x$, $u_y$ and $u_z$), the tractions ($t_x$, $t_y$ and $t_z$) are the unknowns. On the other hand, for elements with prescribed tractions ($t_x$, $t_y$ and $t_z$), the displacements ($u_x$, $u_y$ and $u_z$) are the unknowns. The set of $3T$ algebraic equations may be written in the form

$$[K]_{3T \times 3T}\{U\}_{3T \times 1} = \{F\}_{3T \times 1}$$

(14)

where $\{U\}$ denotes the vector of unknowns, which consists of unknown displacements and unknown tractions. The matrix $[K]$ is fully populated, in general. Solving (14) for $\{U\}$, one can straight away obtain the values of the unknowns, be it unknown displacements or unknown tractions.

Now, the method used to find the integrals of the fundamental solutions over the elements is explained, i.e., the goal now is to evaluate the integrals

$\int_{S^m} U_{xx}(P_e, Q_m) dS^m$ , $\int_{S^m} T_{xx}(P_e, Q_m) dS^m$ ,

$\int_{S^m} U_{xy}(P_e, Q_m) dS^m$ , $\int_{S^m} T_{xy}(P_e, Q_m) dS^m$ ,

$\int_{S^m} U_{xz}(P_e, Q_m) dS^m$ , $\int_{S^m} T_{xz}(P_e, Q_m) dS^m$ ,

$\int_{S^m} U_{yx}(P_e, Q_m) dS^m$ , $\int_{S^m} T_{yx}(P_e, Q_m) dS^m$ ,

$\int_{S^m} U_{yy}(P_e, Q_m) dS^m$ , $\int_{S^m} T_{yy}(P_e, Q_m) dS^m$ ,

$\int_{S^m} U_{yz}(P_e, Q_m) dS^m$ , $\int_{S^m} T_{yz}(P_e, Q_m) dS^m$ ,

$\int_{S^m} U_{zx}(P_e, Q_m) dS^m$ , $\int_{S^m} T_{zx}(P_e, Q_m) dS^m$ ,

$\int_{S^m} U_{zy}(P_e, Q_m) dS^m$ , $\int_{S^m} T_{zy}(P_e, Q_m) dS^m$ ,

$\int_{S^m} U_{zz}(P_e, Q_m) dS^m$ , $\int_{S^m} T_{zz}(P_e, Q_m) dS^m$

These integrals are evaluated by numerical integration, as explained in Chapter 6 of [4]. All these integrals are evaluated by using the common formula

$$\int_{S^m} f(x,y,z) dS^m = \int_0^1 \int_0^{1-v} f(x,y,z) J^m du dv$$

$$= \int_0^1 \int_0^1 f(x,y,z)(1-v) J^m dt dv$$

$$\cong \frac{1}{16} \sum_{k=1}^{16} f(t_k, v_k)$$

$$= \frac{1}{16} \sum_{k=1}^{16} f(x_k, y_k, z_k)(1-v_k) J^m$$

(15)

Equation (15) may simply be written as

$$\int_{S^m} f(x,y,z) dS^m = \frac{1}{16} \sum_{k=1}^{16} f(x_k, y_k, z_k)(1-v_k) J^m$$

(16)

In equation (16), $f(x, y, z)$ is the fundamental solution (i.e., $U_{xx}$, $T_{xx}$, $U_{xy}$, $T_{xy}$, $U_{xz}$, $T_{xz}$, $U_{yx}$, $T_{yx}$, $U_{yy}$, $T_{yy}$, $U_{yz}$, $T_{yz}$, $U_{zx}$, $T_{zx}$, $U_{zy}$, $T_{zy}$, $U_{zz}$, $T_{zz}$) which needs to be integrated over the element that has the element number $m$.

Let $(x_a, y_a, z_a)$, $(x_b, y_b, z_b)$ and $(x_c, y_c, z_c)$ be the coordinates of the vertices which define the triangular element $m$. Of course, the vertices always have to be properly ordered such that the normal vector to $S^m$ points out of the 3D solid under consideration. Then $J^m$ in (16) is given by

$$J^m = 2\sqrt{\sigma^m (\sigma^m - \alpha^m)(\sigma^m - \beta^m)(\sigma^m - \gamma^m)}$$

(17)

where $\sigma^m = \dfrac{\alpha^m + \beta^m + \gamma^m}{2}$

$\alpha^m = \sqrt{(x_a - x_b)^2 + (y_a - y_b)^2 + (z_a - z_b)^2}$

$\beta^m = \sqrt{(x_b - x_c)^2 + (y_b - y_c)^2 + (z_b - z_c)^2}$

$\gamma^m = \sqrt{(x_c - x_a)^2 + (y_c - y_a)^2 + (z_c - z_a)^2}$

To evaluate $x_k$, $y_k$ and $z_k$ in (16), the following values for $(t_k, v_k)$ have to be noted down.

$(t_1, v_1) = \left( \dfrac{1}{4} + \dfrac{1}{4\sqrt{3}}, \dfrac{1}{4} + \dfrac{1}{4\sqrt{3}} \right)$

$(t_2, v_2) = \left( \dfrac{1}{4} + \dfrac{1}{4\sqrt{3}}, \dfrac{1}{4} - \dfrac{1}{4\sqrt{3}} \right)$

$(t_3, v_3) = \left( \dfrac{1}{4} - \dfrac{1}{4\sqrt{3}}, \dfrac{1}{4} + \dfrac{1}{4\sqrt{3}} \right)$

$(t_4, v_4) = \left( \dfrac{1}{4} - \dfrac{1}{4\sqrt{3}}, \dfrac{1}{4} - \dfrac{1}{4\sqrt{3}} \right)$

$(t_5, v_5) = \left( \dfrac{3}{4} + \dfrac{1}{4\sqrt{3}}, \dfrac{1}{4} + \dfrac{1}{4\sqrt{3}} \right)$

$(t_6, v_6) = \left( \dfrac{3}{4} + \dfrac{1}{4\sqrt{3}}, \dfrac{1}{4} - \dfrac{1}{4\sqrt{3}} \right)$

$$(t_7, v_7) = \left(\frac{3}{4} - \frac{1}{4\sqrt{3}}, \frac{1}{4} + \frac{1}{4\sqrt{3}}\right)$$

$$(t_8, v_8) = \left(\frac{3}{4} - \frac{1}{4\sqrt{3}}, \frac{1}{4} - \frac{1}{4\sqrt{3}}\right)$$

$$(t_9, v_9) = \left(\frac{3}{4} + \frac{1}{4\sqrt{3}}, \frac{3}{4} + \frac{1}{4\sqrt{3}}\right)$$

$$(t_{10}, v_{10}) = \left(\frac{3}{4} + \frac{1}{4\sqrt{3}}, \frac{3}{4} - \frac{1}{4\sqrt{3}}\right)$$

$$(t_{11}, v_{11}) = \left(\frac{3}{4} - \frac{1}{4\sqrt{3}}, \frac{3}{4} + \frac{1}{4\sqrt{3}}\right)$$

$$(t_{12}, v_{12}) = \left(\frac{3}{4} - \frac{1}{4\sqrt{3}}, \frac{3}{4} - \frac{1}{4\sqrt{3}}\right)$$

$$(t_{13}, v_{13}) = \left(\frac{1}{4} + \frac{1}{4\sqrt{3}}, \frac{3}{4} + \frac{1}{4\sqrt{3}}\right)$$

$$(t_{14}, v_{14}) = \left(\frac{1}{4} + \frac{1}{4\sqrt{3}}, \frac{3}{4} - \frac{1}{4\sqrt{3}}\right)$$

$$(t_{15}, v_{15}) = \left(\frac{1}{4} - \frac{1}{4\sqrt{3}}, \frac{3}{4} + \frac{1}{4\sqrt{3}}\right)$$

$$(t_{16}, v_{16}) = \left(\frac{1}{4} - \frac{1}{4\sqrt{3}}, \frac{3}{4} - \frac{1}{4\sqrt{3}}\right)$$

(18)

Now, $u_k$ is calculated as

$$u_k = t_k(1 - v_k)$$

(19)

Next, to calculate $x_k$, $y_k$ and $z_k$ in (16), one needs to also calculate the components of the unit normal vector to the element surface $S^m$. Again, assuming that $(x_a, y_a, z_a)$, $(x_b, y_b, z_b)$ and $(x_c, y_c, z_c)$ are the coordinates of the vertices of the triangular element $m$, the components of the unit normal vector in the $x$, $y$ and $z$ direction are given by

$$n_x^m = \frac{(y_b - y_a)(z_c - z_a) - (z_b - z_a)(y_c - y_a)}{d}$$

$$n_y^m = \frac{(z_b - z_a)(x_c - x_a) - (x_b - x_a)(z_c - z_a)}{d}$$

$$n_z^m = \frac{(x_b - x_a)(y_c - y_a) - (y_b - y_a)(x_c - x_a)}{d}$$

(20)

where $d = [((y_b - y_a)(z_c - z_a) - (z_b - z_a)(y_c - y_a))^2$

$+ ((z_b - z_a)(x_c - x_a) - (x_b - x_a)(z_c - z_a))^2$

$+ ((x_b - x_a)(y_c - y_a) - (y_b - y_a)(x_c - x_a))^2]^{1/2}$

Now, depending on the values of $|n_z^m|$ and $|n_y^m|$, $x_k$, $y_k$ and $z_k$ in (16) can be calculated using the appropriate equation from one of the following (21)-(23).

If $|n_z^m| \geq \frac{1}{\sqrt{3}}$

$$x_k = (x_b - x_a)u_k + (x_c - x_a)v_k + x_a$$
$$y_k = (y_b - y_a)u_k + (y_c - y_a)v_k + y_a$$
$$z_k = -(n_z^m)^{-1}[n_x^m(x_k - x_a) + n_y^m(y_k - y_a)] + z_a$$

(21)

Else, if $|n_z^m| < \frac{1}{\sqrt{3}}$ and $|n_y^m| \geq \frac{1}{\sqrt{3}}$

$$x_k = (x_b - x_a)u_k + (x_c - x_a)v_k + x_a$$
$$z_k = (z_b - z_a)u_k + (z_c - z_a)v_k + z_a$$
$$y_k = -(n_y^m)^{-1}[n_x^m(x_k - x_a) + n_z^m(z_k - z_a)] + y_a$$

(22)

Else, if $|n_z^m| < \frac{1}{\sqrt{3}}$ and $|n_y^m| < \frac{1}{\sqrt{3}}$

$$y_k = (y_b - y_a)u_k + (y_c - y_a)v_k + y_a$$
$$z_k = (z_b - z_a)u_k + (z_c - z_a)v_k + z_a$$
$$x_k = -(n_x^m)^{-1}[n_y^m(y_k - y_a) + n_z^m(z_k - z_a)] + x_a$$

(23)

Equations (21)-(23) are also used to evaluate the Cartesian coordinates of $P_e$ or $Q_m$, which may be denoted as $(x^m, y^m, z^m)$, by setting $u_k = \frac{1}{4}$ and $v_k = \frac{1}{2}$. Hence, for element $m$, $(x^m, y^m, z^m)$ which is the chosen point inside the element $m$ and which is the only point on the element where displacement or traction is considered (the other points on the element having the same value of displacement or traction as that of this point), is given by (24)-(26).

If $|n_z^m| \geq \dfrac{1}{\sqrt{3}}$

$$x^m = (x_b - x_a)\frac{1}{4} + (x_c - x_a)\frac{1}{2} + x_a$$

$$y^m = (y_b - y_a)\frac{1}{4} + (y_c - y_a)\frac{1}{2} + y_a$$

$$z^m = -(n_z^m)^{-1}[n_x^m(x^m - x_a) + n_y^m(y^m - y_a)] + z_a$$

(24)

Else, if $|n_z^m| < \dfrac{1}{\sqrt{3}}$ and $|n_y^m| \geq \dfrac{1}{\sqrt{3}}$

$$x^m = (x_b - x_a)\frac{1}{4} + (x_c - x_a)\frac{1}{2} + x_a$$

$$z^m = (z_b - z_a)\frac{1}{4} + (z_c - z_a)\frac{1}{2} + z_a$$

$$y^m = -(n_y^m)^{-1}[n_x^m(x^m - x_a) + n_z^m(z^m - z_a)] + y_a$$

(25)

Else, if $|n_z^m| < \dfrac{1}{\sqrt{3}}$ and $|n_y^m| < \dfrac{1}{\sqrt{3}}$

$$y^m = (y_b - y_a)\frac{1}{4} + (y_c - y_a)\frac{1}{2} + y_a$$

$$z^m = (z_b - z_a)\frac{1}{4} + (z_c - z_a)\frac{1}{2} + z_a$$

$$x^m = -(n_x^m)^{-1}[n_y^m(y^m - y_a) + n_z^m(z^m - z_a)] + x_a$$

(26)

Now, one can see that (16) can now be evaluated if one knows the expressions for the fundamental solutions (i.e., $U_{xx}$, $T_{xx}$, $U_{xy}$, $T_{xy}$, $U_{xz}$, $T_{xz}$, $U_{yx}$, $T_{yx}$, $U_{yy}$, $T_{yy}$, $U_{yz}$, $T_{yz}$, $U_{zx}$, $T_{zx}$, $U_{zy}$, $T_{zy}$, $U_{zz}$, $T_{zz}$). Using (2) and (3), expressions for the fundamental solutions may be written in the expanded form as given by (27) below. In these equations, $(x_1, y_1, z_1)$ denotes the coordinates of the point $P_e$ while $(x_2, y_2, z_2)$ denotes the coordinates of the point $(x_k, y_k, z_k)$.

$$U_{xx} = \frac{C}{r}\left[C_1 + \left(\frac{dr}{dx}\right)^2\right] \qquad U_{yy} = \frac{C}{r}\left[C_1 + \left(\frac{dr}{dy}\right)^2\right]$$

$$U_{zz} = \frac{C}{r}\left[C_1 + \left(\frac{dr}{dz}\right)^2\right]$$

$$U_{xy} = U_{yx} = \left(\frac{C}{r}\right)\left(\frac{dr}{dx}\right)\left(\frac{dr}{dy}\right)$$

$$U_{yz} = U_{zy} = \left(\frac{C}{r}\right)\left(\frac{dr}{dy}\right)\left(\frac{dr}{dz}\right)$$

$$U_{zx} = U_{xz} = \left(\frac{C}{r}\right)\left(\frac{dr}{dz}\right)\left(\frac{dr}{dx}\right)$$

$$T_{xx} = \frac{-C_2}{r^2}\left[\left\{C_3 + 3\left(\frac{dr}{dx}\right)^2\right\}\cos\theta\right]$$

$$T_{yy} = \frac{-C_2}{r^2}\left[\left\{C_3 + 3\left(\frac{dr}{dy}\right)^2\right\}\cos\theta\right]$$

$$T_{zz} = \frac{-C_2}{r^2}\left[\left\{C_3 + 3\left(\frac{dr}{dz}\right)^2\right\}\cos\theta\right]$$

$$T_{xy} = \frac{-C_2}{r^2}\left[3\left(\frac{dr}{dx}\right)\left(\frac{dr}{dy}\right)\cos\theta - C_3\left(n_y^m\frac{dr}{dx} - n_x^m\frac{dr}{dy}\right)\right]$$

$$T_{yx} = \frac{-C_2}{r^2}\left[3\left(\frac{dr}{dy}\right)\left(\frac{dr}{dx}\right)\cos\theta - C_3\left(n_x^m\frac{dr}{dy} - n_y^m\frac{dr}{dx}\right)\right]$$

$$T_{yz} = \frac{-C_2}{r^2}\left[3\left(\frac{dr}{dy}\right)\left(\frac{dr}{dz}\right)\cos\theta - C_3\left(n_z^m\frac{dr}{dy} - n_y^m\frac{dr}{dz}\right)\right]$$

$$T_{zy} = \frac{-C_2}{r^2}\left[3\left(\frac{dr}{dz}\right)\left(\frac{dr}{dy}\right)\cos\theta - C_3\left(n_y^m\frac{dr}{dz} - n_z^m\frac{dr}{dy}\right)\right]$$

$$T_{zx} = \frac{-C_2}{r^2}\left[3\left(\frac{dr}{dz}\right)\left(\frac{dr}{dx}\right)\cos\theta - C_3\left(n_x^m\frac{dr}{dz} - n_z^m\frac{dr}{dx}\right)\right]$$

$$T_{xz} = \frac{-C_2}{r^2}\left[3\left(\frac{dr}{dx}\right)\left(\frac{dr}{dz}\right)\cos\theta - C_3\left(n_z^m\frac{dr}{dx} - n_x^m\frac{dr}{dz}\right)\right]$$

(27)

where $r = \sqrt{(x_2-x_1)^2 + (y_2-y_1)^2 + (z_2-z_1)^2}$

$$\frac{dr}{dx} = \frac{(x_2-x_1)}{r} \qquad \frac{dr}{dy} = \frac{(y_2-y_1)}{r}$$

$$\frac{dr}{dz} = \frac{(z_2-z_1)}{r}$$

$$\cos\theta = \frac{1}{r}\left[(x_2-x_1)n_x^m + (y_2-y_1)n_y^m + (z_2-z_1)n_z^m\right]$$

(From equation (4))

Other notations have the same meanings as earlier

$n_x^m, n_y^m$ and $n_z^m$ are constant over an element $m$

$n_x^m, n_y^m$ and $n_z^m$ are different for different elements, in general

Here one can note that since there are sixteen $(x_k, y_k, z_k)$ for every element $m$, when one integrates a fundamental solution over an element surface $S^m$ (which contains the point $Q_m$), for every $(x_1, y_1, z_1)$, there are sixteen different $(x_2, y_2, z_2)$. Further, when the whole code is considered, since the total number of elements equals $T$, for every $(x_1, y_1, z_1)$, there are $16T$ different $(x_2, y_2, z_2)$; and there are $T$ different $(x_1, y_1, z_1)$ in total.

III. THE CODE EXPLAINED

The present code is explained in this section. The variables in the program (code) may or may not be identical to the corresponding notations in the previous (i.e., 'Theory') section.

One can note that there are eight supplementary files that are available with the online version of the present paper. Logging into the website (after creating an account for free) of the present journal may be necessary to access the supplementary files. The present code is available through either of 'code_medium.m' or 'code_high.m'. The only difference between the files is that they contain different input data; otherwise codes are the same. Since the .m files 'code_medium.m' and 'code_high.m' are self-contained (i.e., since they contain input data also), they may readily be run from within MATLAB (author has used MATLAB R2010b). When the file 'code_medium.m' is run, the result obtained in the MATLAB Command Window is manually saved into 'result_medium.txt'. Similarly, when the file 'code_high.m' is run, the result obtained is manually saved into 'result_high.txt'. To use the present code to solve any other 3D linear elastostatic problem, one need to just change the input data portion of either of 'code_medium.m' or 'code_high.m'.

The file 'mesh_medium.stl' is the .stl file which represents the example 3D geometry discretized into 172 boundary elements. The file 'mesh_high.stl' represents the same geometry with 428 boundary elements. The .stl files are manually edited and formatted in a text editor such as Notepad into the format of the input mesh for the present code 'code_medium.m' or 'code_high.m', and saved as .txt files. The 'mesh_medium.stl' is edited, formatted and then saved as 'mesh_medium.txt' whereas 'mesh_high.stl' is edited, formatted and saved as 'mesh_high.txt'. Since 'code_medium.m' and 'code_high.m' contain input data also, 'code_medium.m' already contains 'mesh_medium.txt' and 'code_high.m' already contains 'mesh_high.txt'. To use the present code to solve problems other than the present test problem, in the similar fashion, one needs to prepare a mesh for the geometry under consideration, and use the prepared mesh as an input data for either of 'code_medium.m' or 'code_high.m', the other input data being the specification of boundary conditions, i.e., the specification of displacements for elements with specified displacements and the specification of tractions for the rest of the elements.

Now, the code 'code_medium.m' is explained in detail, line by line. Except input data portion, 'code_medium.m' and 'code_high.m' are identical. For that matter, except input data portion, the code to solve any other 3D linear elastostatic problem would be the same as 'code_medium.m'.

The 5$^{th}$ line of 'code_medium.m' specifies the modulus of elasticity, while the 6$^{th}$ line specifies the Poisson's ratio. The 7$^{th}$ line specifies the displacement boundary conditions; "161 0 0 0" here means that the element number 161 has specified zero displacements along $x$, $y$ and $z$ directions; similarly, "162 0 0 0" means the element 161 is fixed; same for elements up to 166. The 8$^{th}$ line specifies the nonzero force boundary conditions; "167 0 0 10000" here means that the element 167 is subjected to zero traction along $x$ direction, zero traction along $y$ direction, but 10000 units of traction along the $z$ direction; same is the case for elements up to 172. Now, one can see that the elements which are not subjected to displacement boundary conditions and are not subjected to nonzero force boundary conditions also, are subjected to zero force boundary conditions; Lines 9-12 specify zero force boundary conditions; tractions on the elements mentioned here are zero in $x$, $y$ and $z$ directions. Line 13 combines zero and nonzero force boundary conditions. The variable 'xyzofelements' in line 14 takes a mesh as input; the mesh has 172 elements; the mesh describes the 3D geometry under consideration; mesh is just copy-pasted from 'mesh_medium.txt'; "1  2.000000e+000  0.000000e+000  1.000000e+001;  1  1.000000e+000  0.000000e+000  1.000000e+001;  1  1.000000e+000  0.000000e+000  5.000000e+000" in line 14 means that for element 1, $x_a$ =2.000000e+000, $y_a$ = 0.000000e+000, $z_a$ = 1.000000e+001, again for element 1, $x_b$ = 1.000000e+000, $y_b$ = 0.000000e+000, $z_b$ = 1.000000e+001, again for element 1, $x_c$ = 1.000000e+000, $y_c$ = 0.000000e+000, $z_c$ = 5.000000e+000; lines 15-185 have similar meaning.

Data entered until now form the input portion of the code. The code now contains the geometry, boundary conditions, and the material property. One can use the code 'code_medium.m' to solve any other 3D linear elastostatic problem by just changing this portion of the code to provide the data that are relevant to the new problem.

Lines 186-191 evaluate the constants $G$, $C$, $C_1$, $C_2$, $C_3$ and $n$. Lines 192-193 calculate the total number of elements. Lines 194-195 calculate the total number of elements with displacement boundary condition. Lines 196-197 calculate the total number of elements with force boundary condition. Lines 198-204 are initializations. Lines 205-237, using (20) calculate $n_x^m$, $n_y^m$, $n_z^m$, using (17) calculate $J^m$, using (24) or (25) or (26) calculate $x^m$, $y^m$, $z^m$; these are calculated for each and every element. Lines 238-239 input the values of $t_k$ and $v_k$, as given in (18). Line 240 calculates $u_k$ using (19). Lines 241-242 are initializations.

Purpose of lines 243-520 is to calculate $[K]$ and $\{F\}$ of (14). The outermost *for* loop starts at line 243 and ends at line 520; the iteration here is for different values of $P_e$; hence, there are as many iterations of this loop as the total number of elements. The *for* loop starting at line 244 and ending at line 381 iterates for every element with force boundary condition, for a fixed $P_e$ defined by the outer loop; lines 245-284 evaluate the values of $x_k$, $y_k$, $z_k$ using (21) or (22) or (23) (here, x1 means $x_a$, y2 means $y_b$ etc.); lines 285-302 are just initializations. There is one more *for* loop which starts at line 303 and ends at line 345; the loop is within the previous loop; purpose of this loop is to evaluate (16); of course, the loop evaluates sixteen times the value of the right hand side of (16); lines 304-308 evaluate the values of r, dr/dx, dr/dy, dr/dz and $\cos\theta$, for every $(x_k, y_k, z_k)$ of the element defined by the outer loop; lines 309-317 and lines 327-335 evaluate the expressions in (27) at each of $(x_k, y_k, z_k)$, and lines 318-326 and lines 336-344 add the evaluated values to accumulate to sixteen times the value of the right hand side of (16). Coming out of the innermost *for* loop, lines 346-363 evaluate the right hand side of (16); lines 364-375 build $[K]$ and $\{F\}$ of (14); here one can remember that (14) is the same as (11)-(13) considered together. Lines 376-380 address the case while, during iterations, $e = m$; in this case, terms on the left hand side of (11)-(13) are unknowns and hence belong to $[K]\{U\}$ and hence $[K]$ (not $\{F\}$) has to be modified by adding '0.5' to the appropriate elements of $[K]$, as has been done in lines 376-380; '0.5' here arises out of '1/2' in $\frac{1}{2}u_x$, $\frac{1}{2}u_y$ and $\frac{1}{2}u_z$, on the left hand side of (11)-(13). Now, the *for* loop from the line 382 to the line 520 iterates for every element with force boundary condition, for a fixed $P_e$ defined by the outer loop; here, lines 382-513 have the same purpose as the lines 244-375. Again, lines 514-518 address the case while, during iterations, $e = m$; in this case, terms on the left hand side of (11)-(13) are known and hence belong to $\{F\}$ and hence $\{F\}$ (not $[K]$) has to be modified, as has been done in lines 514-518; again, '0.5' here arises out of '1/2' in $\frac{1}{2}u_x$, $\frac{1}{2}u_y$ and $\frac{1}{2}u_z$, on the left hand side of (11)-(13).

Now, line 522 calculates $\{U\}$ of (14). Line 524 displays calculated values of the unknowns in the MATLAB Command Window; the unknowns could be either displacements $(u_x, u_y, u_z)$ or tractions $(t_x, t_y, t_z)$; for elements with known displacements, tractions are the unknowns; for elements with known tractions, displacements are the unknowns; line 524 displays the result $\{U\}$ in this format: value of the unknown (in the *x* direction, for element number 1), value of the unknown (in the *y* direction, for element number 1), value of the unknown (in the *z* direction, for element number 1), value of the unknown (in the *x* direction, for element number 2), value of the unknown (in the *y* direction, for element number 2), value of the unknown (in the *z* direction, for element number 2), value of the unknown (in the *x* direction, for element number 3) etc.

## IV. ILLUSTRATION AND VERIFICATION

In the present section, the present code is tested by using the code to solve a simple problem with the known solution, i.e., a bar subjected to end force.

Geometry of the test problem is a prismatic bar. The bar has a (4 *mm* x 4 *mm*) cross section, and the bar is 100 *mm* long. One end of the bar is fixed, while the other end is loaded with 160000 *N* force in the axial direction. The coordinates of the vertices which describe the fixed end of the bar are given by (0,4,0), (4,4,0), (4,0,0) and (0,0,0). The coordinates of the vertices which describe the loaded end are given by (0,0,100), (4,0,100), (4,4,100) and (0,4,100). All dimensions here are expressed in millimeters. The problem is to find the displacement at the loaded end upon the application of the load. Modulus of elasticity is assumed as 200000 *N/mm²*, and the Poisson's ratio is assumed to be equal to 0.33.

For simple geometries like a prismatic bar, one can manually prepare a mesh. But, here, since it is a cumbersome and also an error prone process to manually prepare the mesh, the commercial software Rhinoceros (Version 3.0) is used for this purpose. Of course, the mesh may be prepared by using any of the much commercial or free software that can do the job. First, the prismatic bar is constructed in Rhinoceros; then the geometry is saved as a .stl file. A .stl file describes a 3D geometry in terms of a 3D surface mesh consisting of triangles. Rhinoceros has the option to save a 3D geometry as an .stl file, with different total number of triangles, i.e., one can save the geometry in different resolutions. One should remember to save .stl files in the ASCII format; this format is human readable. Here, the geometry constructed in Rhinoceros is saved with two different resolutions, which resulted in a total of 172 and 428 elements. The mesh with 172 elements is named as 'mesh_medium.stl', and the mesh with 428 elements is named as 'mesh_high.stl'. When 'mesh_medium.stl' is opened in Notepad, on the 4th line, one can read this: "vertex 2.000000e+000  0.000000e+000  1.000000e+001". This means that for the first element, $x_a$ = 2.000000e+000, $y_a$= 0.000000e+000, $z_a$=1.000000e+001. The 5th line reads as: "vertex 1.000000e+000  0.000000e+000  1.000000e+001" which means that for the first element, $x_b$ = 1.000000e+000, $y_b$= 0.000000e+000, $z_b$=1.000000e+001. Similarly 6th line means that $x_c$ = 1.000000e+000, $y_c$= 0.000000e+000, $z_c$=5.000000e+000, for the first element. In the same way, lines 11-13 give the coordinates of $x_a$, $y_a$, $z_a$, $x_b$, $y_b$, $z_b$, $x_c$, $y_c$, $z_c$, for the second element; lines 18-20 give the coordinates of these for the third element, and so on. Now, the file 'mesh_medium.stl' is edited and formatted to the form that is saved as 'mesh_medium.txt'. Same way, 'mesh_high.txt' is obtained from 'mesh_high.stl'. Mesh data from 'mesh_medium.txt' or 'mesh_high.txt' can readily be cut-pasted into the present code.

Now, one needs to identify the elements which are fixed, and the elements which are subjected to tractions. For the example problem considered here, one can note that the elements that have the $z$ coordinates of all their vertices equal to zero are the ones which are fixed, i.e., they are the elements that are subjected to displacement boundary conditions, with all the displacements being zero. One can also note that the elements that have the $z$ coordinates of all their vertices equal to 100 are subjected to traction in the $z$ direction. Hence, for the lower resolution mesh, by looking at 'mesh_medium.txt', one can note that the elements 161-166 are fixed, the elements 167-172 are subjected to nonzero tractions, and the other elements (i.e., the elements 1-160) are subjected to zero traction. Similarly, for the higher resolution mesh, by looking at 'mesh_high.txt', one can note that the elements 409-418 are fixed, the elements 419-428 are subjected to nonzero tractions, and the other elements (i.e., the elements 1-408) are subjected to zero traction.

The nonzero traction in the $z$ direction (for the elements on the loaded end) is given by

$$t_z = \frac{Force}{Area} = \frac{160000}{4 \times 4} = 10000 N/mm^2$$

This value is the same whether one uses a medium resolution mesh or a high resolution mesh.

For the test problem considered here, all input data (discussed in the previous four paragraphs) are already contained in the codes 'code_medium.m' and 'code_high.m', for the medium resolution mesh and high resolution mesh cases respectively.

After running the codes 'code_medium.m' and 'code_high.m', results are saved in the files 'result_medium.txt' and 'result_high.txt' respectively.

Considering 'result_medium.txt', the last eighteen rows give the displacement solutions for the last six elements (the last six elements are the ones which are subjected to nonzero tractions). The solutions, as obtained from the last eighteen rows of the file 'result_medium.txt', are tabulated in Table I.

TABLE I. DISPLACEMENT SOLUTIONS AT THE LOADED END (FOR MEDIUM RESOLUTION MESH)

| Element No. | $u_x*10^3$ mm | $u_y*10^3$ mm | $u_z*10^3$ mm |
|---|---|---|---|
| 167 | 0.000475723884518 | 0.000239277914653 | 0.004667990981856 |
| 168 | -0.000288557449754 | 0.000350459191737 | 0.004640913749073 |
| 169 | -0.000227579883863 | -0.000253099680184 | 0.004662295196899 |
| 170 | 0.000350870893377 | -0.000146245484676 | 0.004666785214835 |
| 171 | 0.000061463916240 | 0.000227509412390 | 0.004136294937252 |
| 172 | -0.000001691317047 | -0.000015656433286 | 0.004147373087041 |

Now, considering 'result_high.txt', the last thirty rows give the displacement solutions for the last ten elements (the last ten elements are the ones which are subjected to nonzero tractions). The solutions, as obtained from the last thirty rows of the file 'result_high.txt', are tabulated in Table II.

TABLE II. DISPLACEMENT SOLUTIONS AT THE LOADED END (FOR HIGH RESOLUTION MESH)

| Element No. | $u_x*10^4$ mm | $u_y*10^4$ mm | $u_z*10^4$ mm |
|---|---|---|---|
| 419 | 0.000014972072838 | 0.000069229499481 | 0.000548700220572 |
| 420 | -0.000053933725093 | 0.000032473982250 | 0.000533494879700 |
| 421 | -0.000084785786018 | 0.000055690465021 | 0.000561328889982 |
| 422 | 0.000019080155101 | 0.000032731914270 | 0.000534518762716 |
| 423 | -0.000027613176014 | -0.000011163464531 | 0.000553379281037 |
| 424 | 0.000005898472131 | 0.000062564046813 | 0.000530196484107 |
| 425 | 0.000032708654327 | 0.000036306507440 | 0.000554565304403 |
| 426 | -0.000056504866697 | 0.000061990516729 | 0.000528972628751 |
| 427 | -0.000025959290317 | 0.000052674228475 | 0.000538346574184 |
| 428 | -0.000020851664801 | 0.000043533648613 | 0.000537601022532 |

From Table I and Table II, one can note that displacements in the $x$ and $y$ directions are an order of magnitude less than the displacements in the $z$ direction, for all the elements, in general. This is expected since, for the present example problem, displacements in the $z$ direction should be dominant. In fact, as far as the present test problem is concerned, one is interested in the displacements along the $z$ direction only. Now, considering only the displacements along the $z$ direction and rounding off the decimal values into three digits, and comparing the results with the result from the analytical formula, one can compile the tables Table III and Table IV.

For the present test problem, the analytical solution, i.e., the result from the well known analytical formula is obtained as

$$Displacement = \frac{Force \times Length}{Area \times E}$$

where 'Displacement' implies the displacement of the loaded end in the z direction

'Force' implies the total force applied at the loaded end (= 160000 $N$)

'Length' implies the length of the prism (= 100 *mm*)
'Area' implies the cross sectional area of the prism (= (4 x 4) *mm* = 16 *mm*)
$E$ is the modulus of elasticity (= 200000 $N/mm^2$)

Using the above formula, the analytical result is found to be equal to 5 *mm*, for all the elements on the loaded end, for both medium resolution and high resolution meshes.

TABLE III. COMPARISON OF THE RESULTS FROM THE CODE AND THE ANALYTICAL FORMULA (FOR MEDIUM RESOLUTION MESH)

| Element Number | $u_z$ From the Code (*mm*) | $u_z$ From the Analytical Formula (*mm*) |
|---|---|---|
| 167 | 4.668 | 5.000 |
| 168 | 4.641 | 5.000 |
| 169 | 4.662 | 5.000 |
| 170 | 4.667 | 5.000 |
| 171 | 4.136 | 5.000 |
| 172 | 4.147 | 5.000 |

TABLE IV. COMPARISON OF THE RESULTS FROM THE CODE AND THE ANALYTICAL FORMULA (FOR HIGH RESOLUTION MESH)

| Element Number | $u_z$ From the Code (*mm*) | $u_z$ From the Analytical Formula (*mm*) |
|---|---|---|
| 419 | 5.487 | 5.000 |
| 420 | 5.335 | 5.000 |
| 421 | 5.613 | 5.000 |
| 422 | 5.345 | 5.000 |
| 423 | 5.534 | 5.000 |
| 424 | 5.302 | 5.000 |
| 425 | 5.546 | 5.000 |
| 426 | 5.290 | 5.000 |
| 427 | 5.383 | 5.000 |
| 428 | 5.376 | 5.000 |

From Table 3 and Table 4, one can see that the results from the code are in good agreement with the results from the analytical formula. Thus, one can infer that the present code has performed satisfactorily. However, one can note that there is not much improvement in accuracy, when the total number of elements is increased from 172 (which corresponds to the medium resolution mesh) to 428 (which corresponds to the high resolution mesh). The reasons could be that the analytical formula itself is just an approximate one, and the present code solves the present example problem as *a 3D problem*; also, when the boundary elements are limited in number, it may be difficult to apply the boundary conditions accurately. Further, numerical integration here always uses only sixteen function evaluations per element, which may be insufficient sometimes especially since singularity of the fundamental solutions is not addressed in the present work. Using very large number of elements might improve accuracy, or one has to use linear or quadratic elements for better convergence; to improve accuracy, one may need to do higher number of function evaluations per element during numerical integration; to improve the accuracy further, one may have to properly address the singularities of the fundamental solutions also.

## V. CONCLUDING REMARKS

This work presents a code written in the very simple programming language MATLAB, for three dimensional linear elastostatics, using constant boundary elements. Present work is justified by the fact that, to the best of his knowledge, author of the present work, apart from the codes which might be available in the websites that are companions to some non open access books, is not aware of any open access source code available in the internet that is written in any of the programming languages. The present code is tested by using the code to solve a simple problem with the known solution, i.e., a bar subjected to end force. Result from the code matched well with that obtained from the analytical formula, thus verifying the code. The code may be used to solve three dimensional linear elastostatic problems. Present work could also be an educational aid to those who would like to acquire just a working knowledge of the boundary element method, as applied to three dimensional elastostatics, quickly and easily. Since the code is available for open access, and also since the code is properly documented (documentation includes listing of all the formulae used) through the present paper, present work would also be of help to those who want to modify and/or build upon the present very basic code to suit their requirements.

The present code is applicable to homogeneous and isotropic materials only, and self weight is not taken into account. In this work, only constant boundary elements are considered. Although constant boundary elements can provide adequate accuracy upon fine discretization, whenever greater accuracy is important, linear and quadratic elements may help to get highly accurate results quickly. Since the emphasis in this work is on readability, the code is not optimized for efficiency. Numerical integration here always uses only sixteen function evaluations per element, which may be insufficient sometimes. Also, singularity of the fundamental solutions is not addressed in the present work and hence, while evaluating the integrals, whenever integrand becomes singular, accuracy of the evaluation of the value of the integrals may not be good enough, especially since the present work always uses only sixteen function evaluations per element; this may make the final results less accurate, and even inaccurate sometimes. These are the limitations of the present work.

As future work, singularity of the fundamental solutions has to be properly addressed. Also, there should be a provision in the code to use more number of function evaluations per element, while evaluating the integrals. The code may be improved for better performance, and the code may be parallelized for multiple CPUs/GPUs. The code may also be extended to cover three dimensional nonlinear elasticity. Also, body forces and dynamics may be taken into account. In addition to constant elements, linear and quadratic elements may also be included. The code can further be extended to cover inhomogeneous and anisotropic materials also.

DISCLAIMER

Codes are provided without any guarantee and without any warranty. Author is not responsible for any loss or damage that may arise because of the use of the codes that are made available with this paper.

ACKNOWLEDGMENT

Author is grateful to the Robotics Lab, Department of Mechanical Engineering & Centre for Product Design and Manufacturing, Indian Institute of Science, Bangalore, INDIA, for providing the necessary infrastructure to carry out this work.

REFERENCES

[1] Watson J. O., "Boundary Elements from 1960 to the Present Day," Electronic Journal of Boundary Elements, Vol. 1, No. 1, pp. 34-46, 2003.
[2] http://peili.hut.fi/BEM/
[3] http://www.boundary-element-method.com/
[4] Ang W.T., A Beginner's Course in Boundary Element Methods, Universal Publishers, Boca Raton, USA, 2007.
[5] http://www.ntu.edu.sg/home/mwtang/bem2011.html
[6] http://www.mathworks.com/matlabcentral/fileexchange/16074-bem-code-for-2d-pulsating-cylinder
[7] http://urbana.mie.uc.edu/yliu/Software/
[8] http://www.ifb.tugraz.at/BEM
[9] Beer G., Smith I. and Duenser C., The Boundary Element Method with Programming, SpringerWienNewYork, 2008.
[10] Ang K.C., "Introducing the Boundary Element Method with MATLAB," International Journal of Mathematical Education in Science and Technology, Vol. 39, No. 4, pp. 505-519, 2008.
[11] Kirkup S. and Yazdani J., "A Gentle Introduction to the Boundary Element Method in Matlab/Freemat," http://www.boundary-element-method.com/AR0814BEM.pdf

.